\documentclass[aps,prl,amsmath,amssymb,preprint,showpacs,longbibliography]{revtex4-1}

\usepackage{graphicx}
\usepackage{color} 

\newlength{\largeur}
\newlength{\doublelargeur}

\newcommand{\degree}{\ensuremath{^\circ}}

\newcommand{\PSI}{\affiliation{Swiss Light Source, Paul Scherrer Institut, 5232 PSI-Villigen, Switzerland}}
\newcommand{\RU}{\affiliation{Radboud University Nijmegen, Institute for Molecules and Materials, 6525 AJ Nijmegen, The Netherlands}}
\newcommand{\CST}{\affiliation{College of Science and Technology, Nihon University, 24-1 Narashinodai 7-chome, Funabashi-shi, Chiba 274-8501, Japan}}
\newcommand{\BESSY}{\affiliation{Helmholtz-Zentrum Berlin f\"{u}r Materialien und Energie GmbH, Albert-Einstein-Strasse 15, 12489 Berlin, Germany}}

\begin{document}


\title{Sub-diffraction sub-100 ps all-optical magnetic switching by passive wavefront shaping}



\author{L. Le Guyader}
\email{loic.le\_guyader@helmholtz-berlin.de}
\PSI{}
\BESSY{}

\author{M. Savoini}
\RU{}

\author{S. El Moussaoui}
\author{M. Buzzi}
\PSI{}

\author{A. Tsukamoto}
\author{A. Itoh}
\CST{}

\author{A. Kirilyuk}
\author{Th. Rasing}
\author{A. V. Kimel}
\RU{}

\author{F. Nolting}
\PSI{}


\date{\today}

\begin{abstract}
The recently discovered magnetization reversal driven solely by a
femtosecond laser pulse has been shown to be a promising way to record
information at record breaking speeds. Seeking to improve the recording density
has raised intriguing fundamental question about the feasibility to combine the
ultrafast temporal with sub-wavelength spatial resolution of magnetic
recording. Here we report about the first experimental demonstration of
sub-diffraction and sub-100~ps all-optical magnetic switching. Using
computational methods we reveal the feasibility of sub-diffraction magnetic
switching even for an unfocused incoming laser pulse. This effect is achieved
via structuring the sample such that the laser pulse experiences a passive
wavefront shaping as it couples and propagates inside the magnetic structure.
Time-resolved studies with the help of photo-emission electron microscopy
clearly reveal that the sub-wavelength switching with the help of the passive
wave-front shaping can be pushed into sub-100~ps regime.
\end{abstract}


\maketitle 

The ever increasing demands for faster and denser magnetic recording has been
continuously fueling the search for ways to control magnetization in a medium
by means other than magnetic fields. Several approaches based on excitation by
intense Teraherz pulses,\cite{Tudosa2004, Kampfrath2011, Kubacka2014} electric
fields,\cite{Ohno2000, Kato2004, Lottermoser2004} spin polarized
currents~\cite{Slonczewski1996, Berger1996, Katine2000, Krause2007} or strain
pulses~\cite{Kim2012, Kovalenko2013} have been suggested to control magnetism
at time scale shorter than 100~ps. Femtosecond visible laser pulses have in
particular been shown to offer extensive control from
demagnetization~\cite{Beaurepaire1996} to reversal,\cite{Stanciu2007} over a
large range of materials, from insulators to metals, and from ferro- to
antiferromagnetic orders.\cite{Kirilyuk2010}

Of particular interest for magnetic recording applications is the magnetization
reversal in GdFeCo ferrimagnetic amorphous alloys induced by single femtosecond
laser pulse.\cite{Stanciu2007} While a detailed microscopic understanding of
this all-optical switching (AOS) phenomena is still lacking, it has been shown
that it occurs via the formation of a transient ferromagnetic-like state where
both the rare-earth and the transition metal magnetic moments are aligned
parallel to each other, in strong contrast with the ground-state anti-parallel
alignment.\cite{Radu2011} Total angular moment conservative exchange of spin
moments between the two magnetic sub-lattices has been suggested to explain
this ultrafast counter-intuitive magnetization dynamics~\cite{Mentink2012} and
experimentally observed.\cite{Bergeard2014} Such magnetization dynamics can be
triggered whenever a heat load brings the magnetic sub-lattices out of their
equilibrium with each other.\cite{Ostler2012}

Besides the obvious attractiveness of recording information with ultrashort
femtosecond long excitations, AOS displays numerous interesting features in
view of potential applications. First of all, it has been shown that rare-earth
free based material properties can be engineered to display
AOS.\cite{Evans2014, Mangin2014} Secondly, AOS is an energy efficient process,
with less then 10~fJ of energy necessary to reverse a 20$\times$20~nm$^2$
magnetic domain in GdFeCo.\cite{Savoini2012} Thirdly, a direct write on can be
achieved using circularly polarized laser pulse and taking advantage of the
magnetic circular dichroism of the recording media.\cite{Khorsand2012}
Finally, laser pulses can be focused with plasmonic antenna to spot sizes of
few tens of nanometer.\cite{Stipe2010, Peng2012, Koene2012, Coppens2013}
However, whether sub-diffraction limited sub-100 ps all-optical magnetization
switching is feasible remains to be tested.

Here we report about the first experimental demonstration of sub-diffraction
and sub-100 ps all-optical magnetization switching. Using computational methods
we reveal the feasibility of sub-diffraction magnetic switching even for an
unfocused incoming laser pulse. This effect is achieved via structuring the
sample such that the laser pulse experiences a passive wavefront shaping as it
couples and propagates inside the magnetic structure. Time-resolved studies
with the help of photo-emission electron microscopy clearly reveal that the
sub-wavelength switching with the help of the passive wave-front shaping can be
pushed into sub-100 ps regime.

\begin{figure}[h]
\includegraphics[width=\doublelargeur]{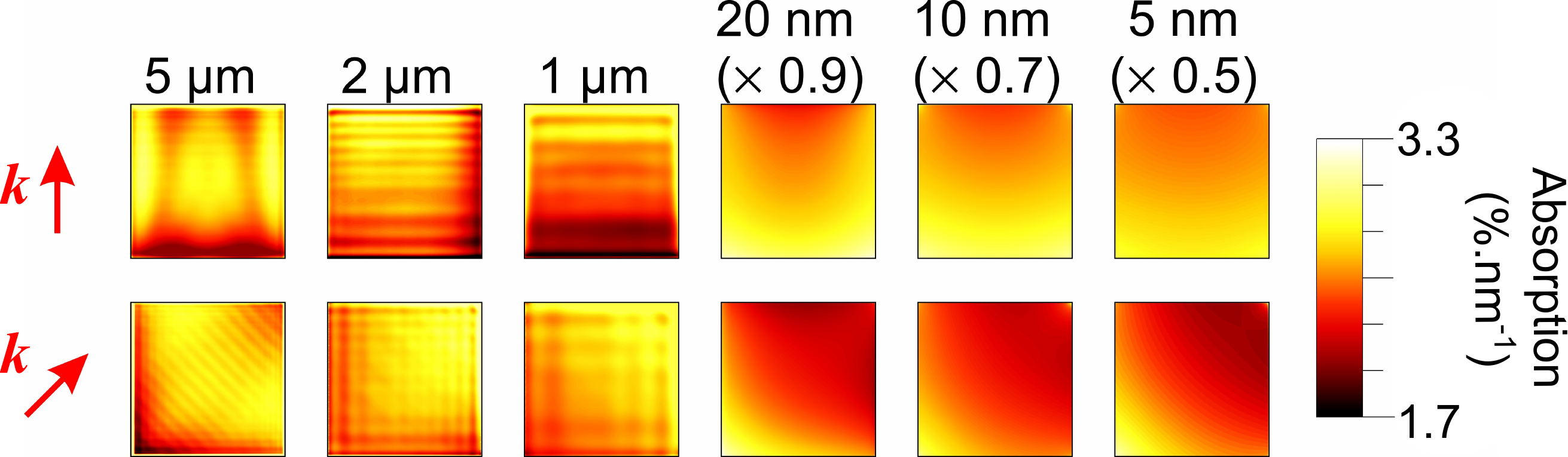}
\caption{\label{fig:FDTDsizes} \textbf{Simulated light absorption as
function of structure sizes and incoming laser direction.} FDTD simulated light
absorption inside structures of different sizes ranging from
5$\times$5~$\mu$m$^2$ down to 5$\times$5~nm$^2$, for two different incoming
azimuthal laser directions indicated by the $k$ left arrow and impinging at
16\degree{} grazing incidence. For the smaller structures, the absorptions
have been reduced by the indicated factor in parentheses such that it falls
into the same range as for the others.}
\end{figure}

In order to demonstrate sub-wavelength all-optical magnetization switching, one
would think of employing near field plasmonic antenna to focus the laser pulse
down to few tens of nm.\cite{Stipe2010, Peng2012, Koene2012, Coppens2013} However, the coupling of the laser
pulse with small structures is a non trivial problem and in some cases, similar
results can be achieve without the use of such plasmonic antennas. We thus
investigated the electromagnetic wave propagation of a femtosecond laser pulse
inside a magnetic structure using finite difference time dependent (FDTD)
simulations. For the modeling, we have chosen a realistic GdFeCo multilayer
structure which is known to display all-optical magnetization switching (AOS),
\textit{i.e.} the ability to reverse permanently its magnetization upon the
sole action of a femtosecond laser pulse.\cite{Stanciu2007} The simulations
were performed for different structure sizes ranging from 5$\times$5~$\mu$m$^2$
down to 5$\times$5~nm$^2$ and for two different incoming azimuthal laser
directions at the same 16\degree{} grazing incidence. The resulting light
absorption profiles are shown in Fig.~\ref{fig:FDTDsizes}. The first striking
feature is that even though the incoming laser pulse is a plane-wave with a
800~nm wavelength, \textit{i.e.} orders of magnitude larger than the smallest
simulated structure, the light absorption inside the structure is inhomogeneous
down to the 5$\times$5~nm$^2$ structure. These absorption
profiles depend on the incoming laser direction, revealing that a particularly
interesting case occurs at 45\degree{} where the absorbed laser energy is
confined within a quarter of the structure. Moreover, these absorption profiles
inhomogeneities are rather strong, displaying a ratio of about 2.0 between the
high and low absorption regions inside the structures down to
20$\times$20~nm$^2$ structure size. This ratio reduces to 50\% for the
10$\times$10~nm$^2$ and 10\% for the 5$\times$5~nm$^2$ structure size. On top
of that, the total absorbed energy increases by a factor of 2.0 from the
largest to the smallest structures, making the smaller structures more
absorbing and thus more energy efficient as previously
reported.\cite{Savoini2012} These focusing and coupling efficiency effects are
the results of the passive wavefront shaping created by the structure's
boundaries and the interference between the waves propagating and absorbed
inside the structure. It must be noted that these effects are not only
present at grazing incidence but also at normal incidence as previously shown
by simulation.\cite{Savoini2012} The grazing incidence geometry offers an
additional degree of freedom such that depending on the orientation of the
boundary with respect to the propagation wave-vector of the light pulse,
different continuity relations take place, resulting in different Fresnel
coefficients.\cite{Pedrotti} This leads, for example, to the intense side lobes
seen in the 5$\times$5~$\mu$m$^2$ at 0\degree{} incoming azimuthal direction
shown in Fig.~\ref{fig:FDTDsizes}.  Refraction, reflection and interferences of
these waves occurring inside the structure, which are best seen in the
45\degree{} incoming direction cases, create strong intensity variations inside
the structures. Strong optical absorption of light leads to the formation of
these features on a small length scale of few nanometers.  These simulations
demonstrate that in the case of an incoming laser pulse at grazing incidence,
it is possible to passively shape the laser pulse wavefront by the structure
geometry such that the absorption is confined into parts of the structure which
are well below the far field diffraction limit. Can this passive wave-front
shaping be employed to all-optically switch a sub-diffraction limited region of
a magnetic structure is the question we address next.

\begin{figure}[h]
\includegraphics[width=\doublelargeur]{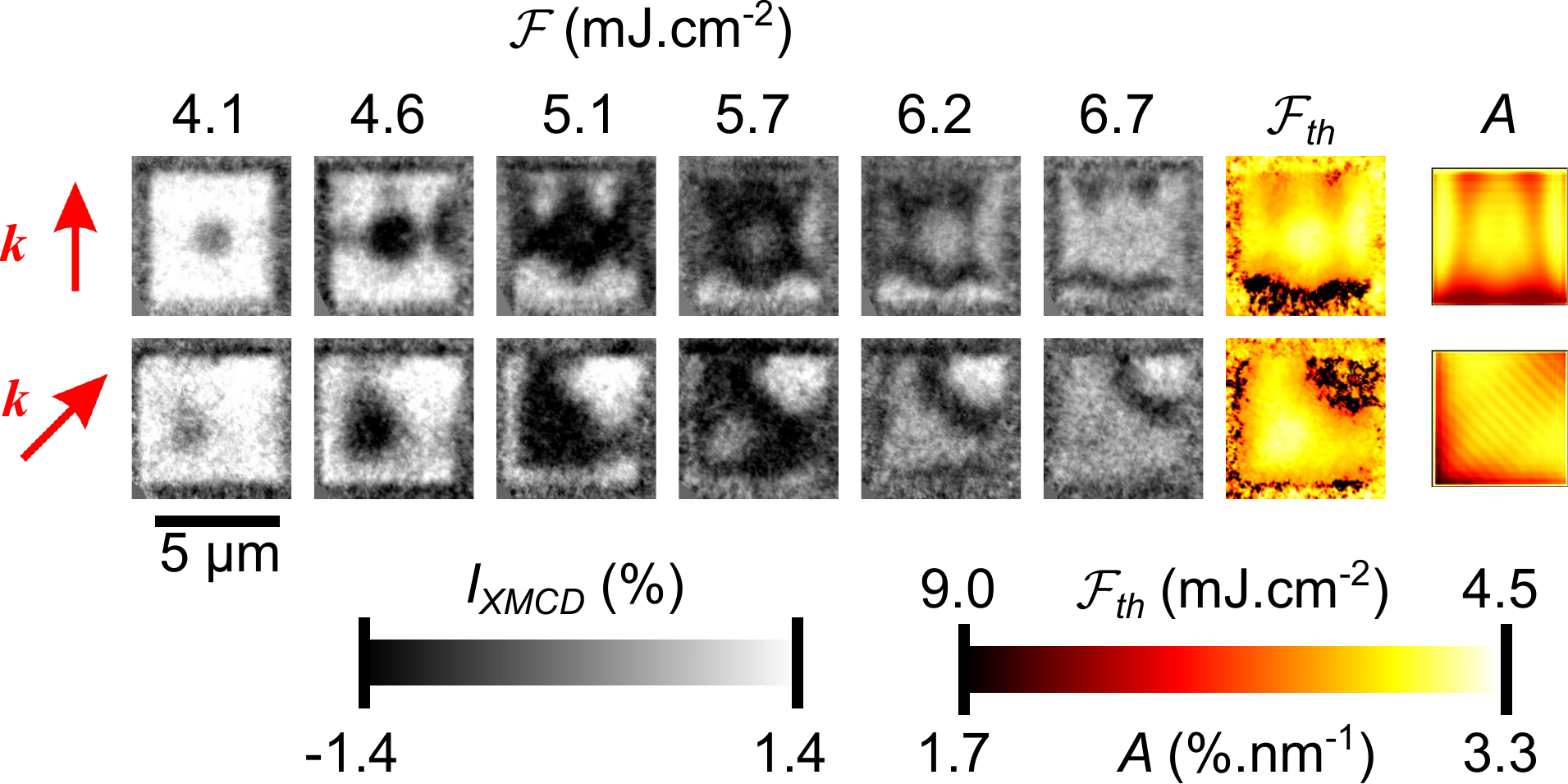}
\caption{\label{fig:fludep} \textbf{Comparison between the experimentally
observed magnetization switching patterns and the simulated light absorption
inside a 5$\times$5~$\mu$m$^2$ GdFeCo structure.} Time-resolved XMCD PEEM
images $I_{XMCD}$ recorded at a fixed time delay of t = 400~ps, as function
of the incoming laser fluence $\mathcal{F}$, as well as the derived AOS
threshold $\mathcal{F}_{th}$ compared with the light absorption $A$ obtained
from the FDTD simulation, for two different azimuthal incoming
laser directions indicated by the $k$ wave-vector.}
\end{figure}

All-optical switching (AOS) occurs via the energy absorbed from the
laser pulse~\cite{Ostler2012} and displays a switching threshold
behavior.\cite{Khorsand2012} This means that below a certain laser fluence, or
better stated, below a certain absorbed energy density, only partial
demagnetization occurs and the sample magnetization recovers to its initial
state. Above this threshold fluence, deterministic magnetization switching
occurs. At even higher fluence, the magnetization switching disappears and
randomly oriented domains are created with no relation to the initial state.
Thus, by investigating the spatially resolved magnetization state in GdFeCo
structures after laser pulse excitation, as function of the laser fluence, it
is possible to study the passive wavefront shaping and focusing experienced by
the laser pulse interacting with the structure and determine whether partial
magnetization switching of the structure is feasible.  However, due to the low
coercivity of the GdFeCo alloys, the switched domains are likely to reorganize
after switching on the relevant length scale here of few hundreds of nanometer.
It is therefor necessary for the sample investigated to probe the magnetization shortly after the laser
pulse. For this, time-resolved X-ray magnetic circular dichroism (XMCD)
photo-emission electron microscopy (PEEM) imaging was employed, which offers
magnetic domain imaging with 70~ps time resolution and 100~nm spatial
resolution. By fixing the time delay $t$ between the laser pump and the x-ray
probe, the spatially resolved intermediate magnetization state inside a
structure at that specific time delay can be recorded.  The XMCD images for a
5$\times$5~$\mu$m$^2$ square microstructure at $t$ = 400~ps after the laser
pulse are shown in Fig.~\ref{fig:fludep}, as function of the incoming laser
fluence, and for two different incoming laser directions indicated by the
arrows (top and bottom row). It is first important to note that at this
relatively long time delay of a few hundred picosecond, both the Gd and the
FeCo sub-lattice magnetization are again in equilibrium with each other such
that measuring only one sub-lattice is enough to characterize the sample
magnetization.\cite{Radu2011} This time scale is on the other hand short enough
to probe the transient longitudinal magnetization dynamics occurring, in
particular whether partial or total demagnetization or magnetization switching
has taken place. The initial state of this microstructure is mono-domain due to
a static applied magnetic field of 50~mT and display a homogeneous white XMCD
contrast. For the low fluence case of $\mathcal{F}$ = 4.1~mJ.cm$^{-2}$, one can
see that the contrast is not uniform anymore, and that some partial
demagnetization has occurred at the center of the structure. As the fluence is
increased, the intermediate magnetization state changes, and each area within
the microstructure varies from a partial demagnetization (white to grey XMCD
contrast) to a full demagnetization (grey XMCD contrast) to a reversed
magnetization (black contrast) and back to a demagnetization state again. This
is essentially the known phase diagram of AOS with its threshold behavior as
explained before.\cite{Khorsand2012} From these series of XMCD images, it is
possible to estimate at which laser fluence $\mathcal{F}_{th}$ each region
inside the structure switches.

The fact that the resulting $\mathcal{F}_{th}$ pattern completely changes with
the incoming laser direction rules out an intrinsic inhomogeneous
$\mathcal{F}_{th}$ due for example to chemical inhomogeneities in the structure
composition whose effects have been seen in other studies.\cite{graves2013} It
is thus not the fluence threshold $\mathcal{F}_{th}$ that varies inside the
structure but the light absorption that does, even though the laser spot size
of 30~$\mu$m$\times$105~$\mu$m$^2$ is much larger than the structure itself.
The spatially resolved FDTD simulated light absorption $A$ inside the
5$\times$5~$\mu$m$^2$ structure are shown in Fig.~\ref{fig:fludep} as well. A
very good qualitative agreement is obtained, as a region with a low
$\mathcal{F}_{th}$ corresponds as expected to a region with high absorption and
vice versa. The agreement is even reasonably quantitative since the ratio
between the high and low fluence threshold $\mathcal{F}_{th}$ is about a factor
of 2.0, in accordance with the ratio between high and low absorption. It seems
thus clear that focusing by passive wavefront shaping occurs in this
microstructure, and in turns creates an inhomogeneous absorption profile which
leads to a spatially selective AOS.

\begin{figure}[h]
\includegraphics[width=\largeur]{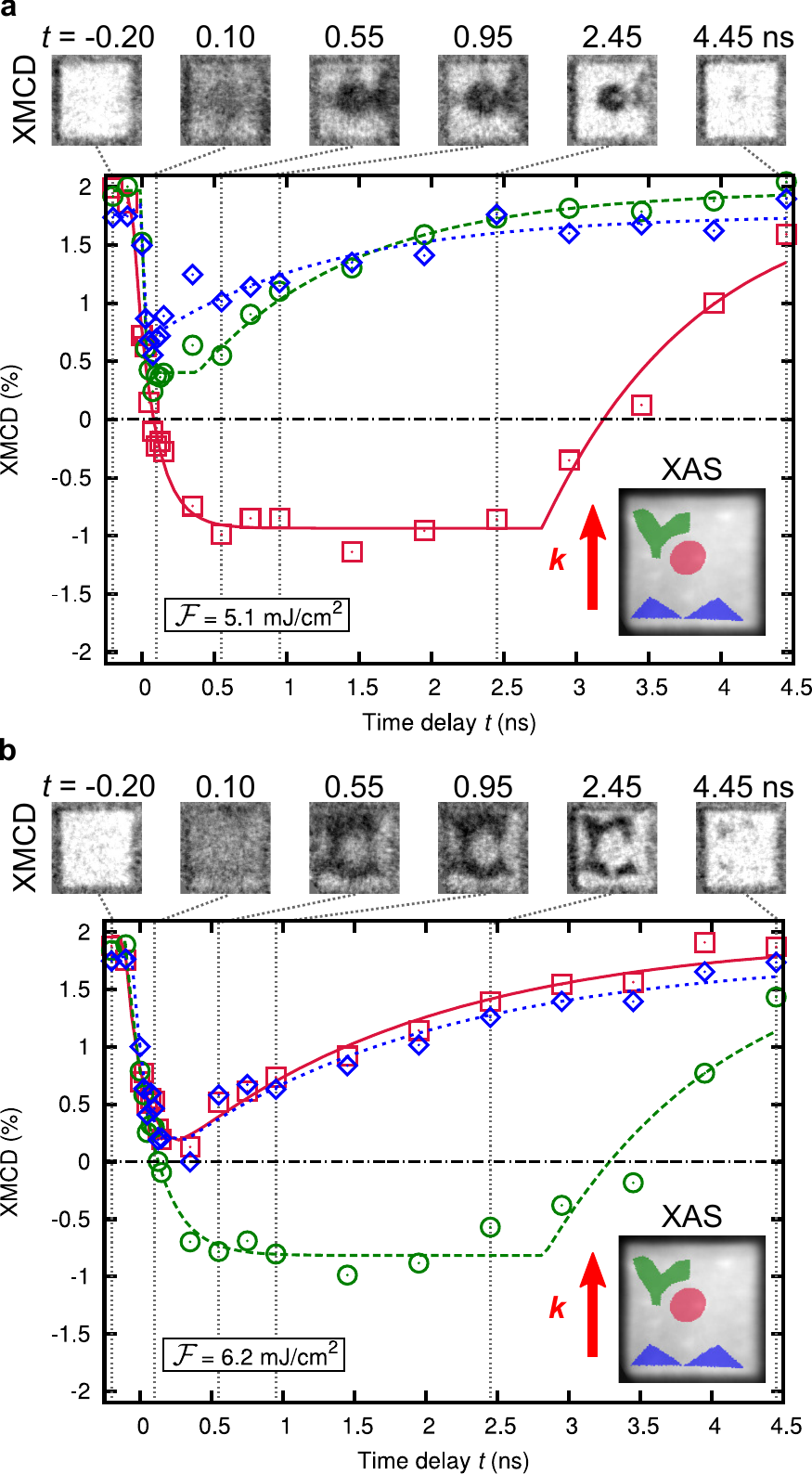}
\caption{\label{fig:5mkmdyn} \textbf{Magnetization dynamics inside a
5$\times$5~$\mu$m$^2$ structure.} Time-resolved XMCD PEEM imaging of the
magnetization dynamics for three different ROI inside the 5$\times$5~$\mu$m$^2$
structures and for two different fluence of (a) $\mathcal{F}$ = 5.1~mJ.cm$^{-2}$
and (b) $\mathcal{F}$ = 6.2~mJ.cm$^{-2}$. The three defined ROI are shown as
colored region in the XAS image in the inset together with the incoming laser
direction $k$. In both cases, a selection of XMCD snapshots at fix time delay
is shown on top of the graph. The lines are a fit to the data with a single
exponential for the demagnetization and remagnetization each.}
\end{figure}

Further proof of this spatially selective AOS can be obtained by looking at
the magnetization dynamics of different regions inside the structures.  The
spatially resolved magnetization dynamics inside a 5~$\mu$m wide microstructure
recorded with time-resolved XMCD PEEM are shown for two different incoming
laser fluences of $\mathcal{F}$ = 5.1~mJ.cm$^{-2}$ in Fig.~\ref{fig:5mkmdyn}(a)
and of $\mathcal{F}$ = 6.2~mJ.cm$^{-2}$ in Fig.~\ref{fig:5mkmdyn}(b). Three
different regions of interest (ROI) have been defined, and for each, the time
dependent magnetic contrast is extracted. The first ROI corresponds to a low
fluence threshold $\mathcal{F}_{th}$, \textit{i.e.} high absorption, the second
ROI to an intermediate case, and the third to a high threshold, \textit{i.e.}
low absorption. In the case of the low incoming fluence shown in
Fig.~\ref{fig:5mkmdyn}(a), only the first ROI shows a magnetization switching
while both other regions show only a partial demagnetization. As the fluence is
increased, as seen in Fig.~\ref{fig:5mkmdyn}(b), the picture drastically
changes. Now, only the second ROI switches, while the first ROI which was
switching at lower fluence now only displays a demagnetization followed by a
recovery to the initial state upon cooling under the action of the
applied magnetic field.

\begin{figure}[h]
\includegraphics[width=\largeur]{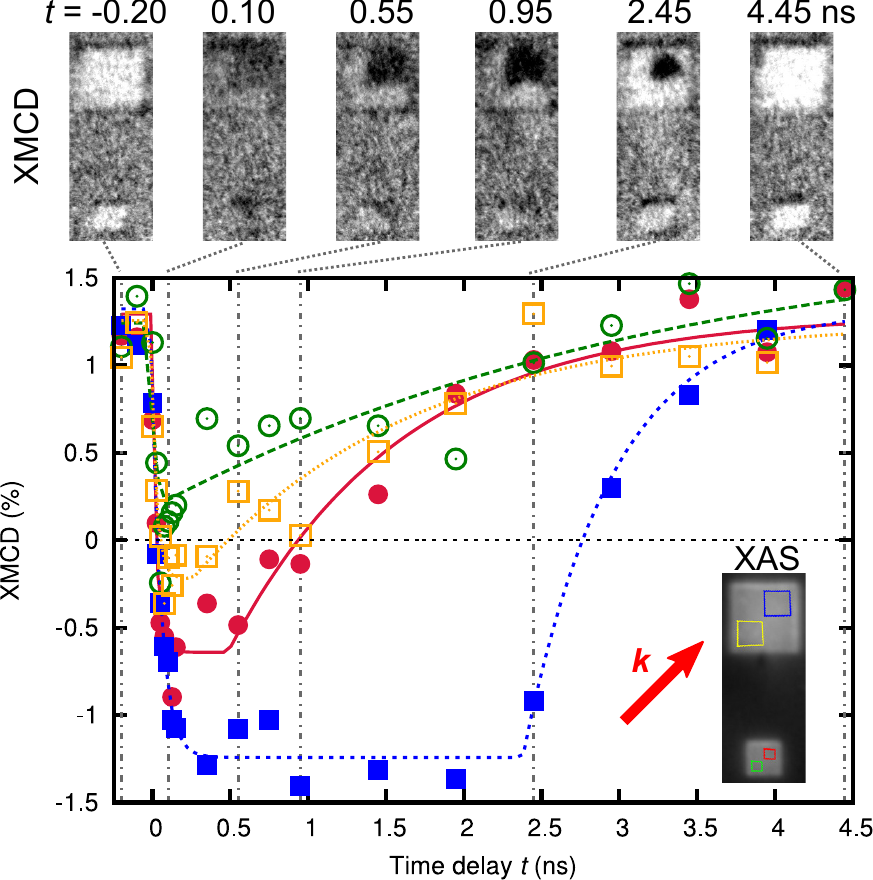}
\caption{\label{fig:tr1m} \textbf{Magnetization dynamics inside a
2$\times$2~$\mu$m$^2$ and 1$\times$1~$\mu$m$^2$ structures.} Time-resolved XMCD
PEEM imaging of the magnetization dynamics for a laser fluence of $\mathcal{F}$
= 5.1~mJ.cm$^{-2}$, for four different ROIs, two of which are inside the
2$\times$2~$\mu$m$^2$ structure and the two others are inside the
1$\times$1~$\mu$m$^2$ structure. The ROI definitions are shown in the XAS image
in the inset together with the 45\degree{} incoming laser direction $k$. A
selection of XMCD snapshots at fix time delay is shown on top of the graph. The
lines are a fit to the data with a single exponential for the demagnetization
and remagnetization each.}
\end{figure}

The same experiments were repeated in 2$\times$2 and 1$\times$1~$\mu$m$^2$
structures and the recorded magnetization dynamics are shown in
Fig.~\ref{fig:tr1m}. In these structures, the switching pattern is different
than for the case of the larger 5$\times$5~$\mu$m$^2$ structures because the
different sizes lead to a different intensity pattern as shown by the FDTD
simulations in Fig.~\ref{fig:FDTDsizes}. In both the 2$\times$2 and
1$\times$1~$\mu$m$^2$ structures, we observe simultaneously a region displaying
AOS and a region displaying demagnetization followed by a quick recovery to the
initial state. In the 1$\times$1~$\mu$m$^2$ structure in particular, the region
showing AOS is about 300~nm wide, which is interestingly smaller than the far
field diffraction limit at this wavelength. This can be understood considering
that the wavelength of the light propagating inside the GdFeCo layer is about 4
times smaller than in vacuum due to the GdFeCo refractive index. Experiments on
smaller structures conducted at the same time were limited by the photo-emitted
electron counts statistic and the instrument spatial resolution in this
time-resolved mode. Nevertheless, the agreement between the FDTD simulations
of the light absorption profile and the experimentally measured spatially
resolved magnetization dynamics is excellent down to a 1$\times$1~$\mu$m$^2$
structure and a 300~nm wide reversed magnetic domain, demonstrating the
spatially selective AOS.

Thus, passive wavefront shaping performed by the structure results in
inhomogeneously absorbed laser energy and is demonstrated experimentally in
GdFeCo structures. It allows for a selective all-optical magnetization
switching inside the microstructure, even though the incoming laser pulse is
homogeneously illuminating the structure. Eventually, at the longer time scale
of a few nanoseconds, the whole structure relaxes to the initial state due to
the applied magnetic field. While this is a requirement here for these
stroboscopic pump-probe experiments in GdFeCo samples, it is not a requirement
to observe this spatially selective switching. In the case of a RE-TM alloy
with a much higher magnetic anisotropy, like for example TbFeCo, the reversed
pattern induced by the laser pulse inside the microstructure would be
stable.\cite{Finazzi2013}


\textbf{Conclusions}\newline{}
In conclusion, we have demonstrated, using FDTD simulations, that passive
wave-front shaping of the laser pulse by the structures shape allows
sub-diffraction focusing of the absorbed energy inside it. We have
experimentally confirmed using time-resolved XMCD PEEM imaging
that this allows sub-diffraction all-optical magnetization switching
of part of the GdFeCo structures. These results open
novel opportunities for very high density data storage media, for example by
either recording several bits of information in a single magnetic bit or by
improving the coupling efficiency between the laser pulse and the magnetic
structure.

\textbf{Methods}\newline{}
\textbf{FDTD simulations.} The electromagnetic wave propagation inside the
structures was simulated with a finite-difference time domain
method.\cite{Lumerical} Various micro- and nano-structures with squared and
circular shape were simulated. They consisted of multilayered-structures
ranging from 5$\times$5~$\mu$m$^2$ down to 5$\times$5~nm$^2$. A variable three
dimensional discretization mesh was used as a function of the pattern size,
ranging from 15$\times$15$\times$1~nm$^3$ for the largest structure down to
0.1$\times$0.1$\times$1~nm$^3$ for the smallest. We considered a plane wave
illumination at a wavelength $\lambda$ = 800~nm, linearly p-polarized,
impinging on the sample with an angle of 16\degree{} grazing incidence
(74\degree{} from the normal). These settings are chosen to correspond to the
experimental conditions, where a Gaussian profile beam is used as illumination
having a FWHM much larger than the structures size. We also used a Gaussian
profile with dimensions comparable with the experimental ones, without
observing substantial differences with the simulations performed with
plane-wave illumination. The structures are sitting on a silicon substrate with
a complex index of refraction \textit{\~{n}}$_\text{Si}$ = n + \textit{i}k =
3.692 + 0.0065\textit{i},\cite{Palik} while the upper half space is vacuum. The
structures are composed of several layers, namely, starting from the bottom
one, AlTi(10~nm)/Si$_3$N$_4$(5~nm)/GdFeCo(20~nm)/Si$_3$N$_4$(3~nm). The
refractive indexes of the layers are \textit{\~{n}}$_\text{AlTi}$ = 2.81 +
5.89\textit{i},\footnote{Calculated after 50\% of Al and 50\% of Ti with the
corresponding index of refraction from Ref.~\onlinecite{Palik}}
\textit{\~{n}}$_\text{Si$_3$N$_4$}$ = 2.0~\cite{Palik} and
\textit{\~{n}}$_\text{GdFeCo}$ = 3.7 + 3.856\textit{i}.\cite{Koene2012} The
light absorption $A = \frac{4\pi nk}{\lambda}|$\textbf{E}$|^2$ where \textbf{E}
is the light electric field is mapped at the center of the GdFeCo layer. A good
convergence of the simulations was obtained with variable time steps smaller
than 0.1~fs and a total simulation time of about 100~fs while the
Fourier-transform limited laser pulse was about 10~fs long.

\textbf{Sample preparation and microstructuring.} The sample consisted of a
multilayer thin-film of composition AlTi(10~nm)\slash Si$_3$N$_4$(5~nm)\slash
Gd$_{24}$Fe$_{66.5}$Co$_{9.5}$(20~nm)\slash Si$_3$N$_4$(3~nm)
grown by magnetron sputtering on a silicon substrate and are essentially the
same as in Ref.~\onlinecite{LeGuyader2012APL}. The structuring of these samples
in squares and discs with sizes ranging from 5$\times$5~$\mu$m$^2$ down to
1$\times$1~$\mu$m$^2$ has been realized via electron beam lithography in
combination with a lift-off process, in which a polymethylmethacrylate resist
is first patterned with an electron beam writer on a Si substrate. This pattern
is then transferred via lift-off after deposition by magnetron sputtering of
the magnetic multilayer AlTi(10~nm)\slash Si$_3$N$_4$(5~nm)\slash
Gd$_{24}$Fe$_{66.5}$Co$_{9.5}$(20~nm)\slash Si$_3$N$_4$(3~nm), resulting in
isolated magnetic structures.\cite{LeGuyader2012JPN} Unstructured areas of
several 100~$\mu$m, quasi-continuous films, and arrays of squares and disks
down to 100~nm were fabricated onto the same sample. In the manuscript, we
focus only on the 5, 2 and 1~$\mu$m squares, since simulations for larger
structures are too time consuming and the signal over noise ratio for smaller
structures is too small.

\textbf{Time-resolved XMCD PEEM measurements}. Spatially resolved images of the
magnetic domain states in these microstructures were obtained with the Elmitec
PEEM at the Surface/Interface: Microscopy (SIM) beamline~\cite{flechsig:319} at
the Swiss Light Source using the XMCD effect at the Fe L$_3$-edge at 708~eV as
a magnetic contrast mechanism. An XMCD asymmetry image is obtained by taking
two total electron yield images measured with opposite x-ray helicities at
resonant energies. The resulting contrast is proportional to the scalar product
of the local magnetization and the incoming X-ray wave
vectors,\cite{Scholl2002} that is, the more parallel the magnetization is to
the x-ray wave vector, the brighter the contrast.  Time-resolved measurements
were performed by taking advantage of the pulsed nature of the X-rays produced
by the SLS synchrotron via the gating of the detection in synchronization to an
isolated x-ray pulse present in the gap of the filling pattern of the storage
ring. This scheme, presented in details in Ref.~\onlinecite{LeGuyader2012JES},
allows stroboscopic pump-probe imaging of the sample with a time resolution
determined by the 70~ps full width at half maximum (FWHM) temporal X-ray pulse
length. In order to perform stroboscopic measurements, the magnetic state of
the sample is recovered after each pump event thanks to a permanent magnet
mounted right underneath the sample and saturating it with a magnetic field of
50~mT. The laser used for the pump is produced by an XL-500 oscillator from
Femtolasers Produktions GmbH and characterized by a $\tau$ = 50~fs laser pulse
length at $\lambda$ = 800~nm wavelength with 500~nJ per pulse at a 5.2 MHz
repetition rate. The laser is then focused on the sample at a grazing incidence
of 16\degree{} to a spot size of FWHM = 30~$\mu$m$\times$105~$\mu$m$^2$
(V$\times$H). Finally, the sample can be azimuthally rotated in situ to perform
experiments with different incoming laser direction.


%

\textbf{Acknowledgments}\newline{}
This work was supported by the European Community's Seventh Framework
Programme FP7/2007-2013 (grants NMP3-SL-2008-214469 (UltraMagnetron),
FP7-NMP-2011-SMALL-281043 (FEMTOSPIN) and 214810 (FANTOMAS)), the European
Research Council ERC Grant agreement No. 257280 (Femtomagnetism), the
Foundation for Fundamental Research on Matter (FOM) and the Technology
Foundation (STW) as well as the Netherlands Organization for Scientific
Research(NWO). Part of this work was performed at the Swiss Light Source,
Paul Scherrer Institut, Villigen, Switzerland. We thank J. Honegger for his
technical support and A. Weber for her support with the nanofabrication.

\textbf{Author contributions}\newline{}
A.V.K., A.K., T.R., and F.N. coordinated the project.
The measurements were performed by L.L.G, S.E.M., M.B. and M.S.
Sample growth and optimization were made by A.T. and A.I. The simulations
were performed by M.S. All the authors contributed to the writing of the
manuscript.

\textbf{Additional information}\newline{}
The authors declare no competing financial interests.

\newpage{}

\end{document}